# Discretization of Temporal Data: A Survey

P. Chaudhari[*1], D. P. Rana[*2], R. G. Mehta[*3], N. J. Mistry[*], M. M. Raghuwanshi[#]

[*]SVNIT, Surat, India
[1]pratikshachaudhari29@gmail.com
[2]dpr@coed.svnit.ac.in
[3]rgm@coed.svnit.ac.in

[#]RGCET, Nagpur, India

*Abstract*—**In real world, the huge amount of temporal data is to be processed in many application areas such as scientific, financial, network monitoring, sensor data analysis. Data mining techniques are primarily oriented to handle discrete features. In the case of temporal data the time plays an important role on the characteristics of data. To consider this effect, the data discretization techniques have to consider the time while processing to resolve the issue by finding the intervals of data which are more concise and precise with respect to time. Here, this research is reviewing different data discretization techniques used in temporal data applications according to the inclusion or exclusion of: class label, temporal order of the data and handling of stream data to open the research direction for temporal data discretization to improve the performance of data mining technique.**

## I. Introduction

The huge amount of data is to be processed in many application areas such as scientific, financial, network monitoring, sensor data analysis [1], [2]. Data mining is important to analyse huge amount of data and with time oriented data, data analysis can be done better and more naturally. Recently, the increasing usages of temporal data in various applications have got focus on research in the field of data mining. Temporal data can be easily obtained and available from scientific and financial applications like ECG, diabetes, daily weather data, sales information and stocks information. A time series is a collection of observations made chronologically. The characteristics of time series data includes: huge in size, high dimensions and updated continuously.

Moreover temporal data have characteristics of numeric and continuous nature, which has to be always considered as a group instead of individual field. There are number of research is going on temporal data, to find similarity, to segment and to search, to reduce in dimension. Temporal discretization refers to the discretization of time series, as a preprocessing step in transforming the temporal data into timely intervals. An effective discretization method not only reduces the dimensionality of data and improve the efficiency of data mining and machine learning algorithm, but also make the knowledge extracted from the discretized dataset more compact, easy to understand and useful. Many data mining algorithms can benefit from a discrete representation of the original data set as the numbers of values for a feature are many, but after the discretization, the number of discrete values is less compare to original number of values [1], [2].

Discretization has a number of advantages: Discrete features reduce memory usage and thus increase representation of the knowledge as data is simplified to understand and with this application of mining technique or knowledge retrieval methods become faster and perfect [2].

There are several ways by which discretization methods can be classified: Splitting versus Merging, Global versus Local, Supervised versus Unsupervised, Static versus Dynamic, and non-Incremental versus Incremental [2], [3], [4], [5].

- The Splitting methods, is a top-down approach of discretization in which one start with an empty set of cut points and gradually divide the interval and subintervals to obtain the discretization. In contrast, the merging method is a bottom-up approach in which it considers all possible cut points and then eliminate these cut points by merging intervals.
- Local methods produce partitions that are applied to localized regions of instance space. Global methods, on the other hand, use the entire instance space and form a mesh over the entire n-dimensional continuous instance space, where each feature is partitioned into regions independent of other attributes.
- Unsupervised methods carry out discretization without the knowledge of class label, whereas the supervised methods utilize the class information to carry out the discretization.
- Static discretization methods require some parameter indicating the maximum number of desired intervals to discretize a feature. Dynamic methods conduct a search through the space of possible k values for all features simultaneously.
- Non-incremental methods consider only the available historical data values. But, with time, the data values that can be completely new are not considered. While in incremental methods they try to cover the new values also.

Here, this research is discussing the different data discretization methods used for different temporal data applications that may have a) class label or no class labels, b) nature of continuous data or static data and c) considering the temporal order or not considering the temporal order.

The organization of rest of this research is as follows: Section 2 discusses the literature review of temporal static data discretization methods which is not considering the temporal order of the timely data and the other discretization methods which is considering the temporal order of the data under the unsupervised and supervised discretization category followed by the comparative analysis of these discretization methods for defined parameters. Final Section 3 summarizes the research for the future scope.

## II. DISCRETIZATION

Many data mining algorithms and tasks can benefit from a discrete representation of the original data set. Discrete representation is more comprehensive to human and can simplify, reduce computational costs and improve accuracy of many algorithms. Discretization is the process of transforming continuous space valued series $X=\{x1,x2,...,xn\}$ into a discrete valued series $Y=\{y1,y2,...,yn\}$. Discretization can be performed recursively on an attribute. The main part of the discretization process is choosing the best *cut points* which split the continuous value range into discrete number of bins usually referred to as *states*.

From the literature review, it is found that most of discretization methods which are used can be majorly categorized as unsupervised and supervised discretization. Here, this section is discussing discretization methods according to this category.

### A. Unsupervised Discretization

When in dataset class information is not available for time series, unsupervised methods are needed. Two common methods used in most of the applications are Equal Width Discretization (EWD) and Equal Frequency Discretization (EFD) [1]. Other than these are K-means Clustering [1], [3] SAX [6], Frequency Dynamic Interval Class (FDIC) [7], where methods: EWD, EFD and K-means clustering are static methods which are not considering temporal order of data.

*1) Equal Width Discretization (EWD):* EWD is a simplest discretization method that divides the range of observed values for a feature into k equal sized bins, where k is a parameter provided by the user [4]. The process involves finding values as the minimum (Vmin) and maximum (Vmax). The interval is computed by dividing the range of observed values for the variable into k number of equally sized bins using the formula Interval=(Vmax-Vmin)/k, where k is a parameter supplied by the user and Boundaries = Vmin+( i * interval) for the i = 1...k-1 boundaries [3]. However, this method of discretization is sensitive to outliers that may drastically skew the range.

The limitations of this static, unsupervised and very simple method are: a) It is the parametric method as the required number of interval is needed from the user. b) It follows the characteristics of how data values are distributed as some intervals may contain much more data points than other and can produce overlapping.

*2) Equal Frequency Discretization:* The equal-frequency algorithm determines the minimum and maximum values of the discretized attribute of n values, it sorts all values in increasing frequency order and then divides the sorted data values into *k* bins such that each interval approximately contains *n/k* data values with adjacent internal values. For equal frequency, many occurrences of a continuous value could cause the occurrences to be assigned into different bins that cause the problem of overlapping.

This method is advantageous and tries to overcome the limitations of the equal-width interval discretization by dividing the domain in intervals with the same distribution of data points. The problems with this static, unsupervised and parametric method is that it is not always possible to generate exactly k equal frequency intervals because it tries to place the data instance with identical value in the same interval.

The author Chaves has applied this method for image histogram discretization [8].

*3) K-means Clustering:* The K-means clustering method is widely used for temporal data discretization in number of applications as it helps to find natural groups and one of the data mining techniques [2], [3], [5]. The author Salvador used the K-means clustering to identify the number of states in a time series dynamically [9], the author Z. Liang et al. used this method to partition the values of the attributes like temperature, salinity, pH, etc. to detect the correlation between environmental factors and ecological events [10].

In K-means clustering, a Euclidean distance measure is a simple distance function used to cluster the data into *k* clusters which are represented by the centroids. The clustering algorithm begins with a random or more educated choice (more efficient due to the sensitivity of the clustering process to the initial selection) of clusters centroids. The second step is to assign each data point to the cluster that has the closest centroid. After every data point has been assigned, the *k* centroids are recalculated as the mean value of each cluster. Two steps are repeated until no data point is reassigned or the *k* centroids no longer change. The resulting clusters or centroids are used as the states of the discretization process.

This static method is advantageous than EWD and EFD methods as the data are grouped together as per their characteristics. But this can be the limitations also and other than this are: a) It is the parametric method, as the required number of interval value k is to be provided by the user. b) The number of intervals of discretization is dependent upon the given k value and seed value of cluster.

*4) Symbolic Aggregate Approximation (SAX):* SAX is the advanced method especially designed for temporal data. The author Keogh et al. applied SAX discretization for improvement of pattern finding performance [11].

The SAX representation of a time series is given by the author Lin et al. in 2003 [6] to convert the piecewise aggregate approximation to symbols. The authors has achieved this by dividing the vertical y-axis into equi probable parts, which replaced by symbol. To do so, parameters like subsequence length and the number of symbols. SAX helps to reduce the dimensionality and lower bounds the distance between any two vectors in the SAX representation is smaller than, or equal to, the distance between these two vectors in the original space.

Before discretization, the author has removed distortions by normalizing each time series and that can follow the Gaussian distribution. Then for the given number of segmentation, transform time series into partial aggregate approximation. And now breakpoints are decided so that they produce equal-sized areas under the Gaussian curve using the breakpoint lookup table that helps to divide the amplitude values of the time series into required number of equi-probable regions. Then depending upon the breakpoint lying area symbols are provided to generate the complete symbol string.

The advantages of this method are a) It is reducing dimensionality depending upon alphabet size. b) It is lower bounding the distance between any two vectors. The SAX method is suitable for most classic data mining tasks like classification and clustering.

The extended-SAX is proposed by the same author in 2007 [12] which included the min and max value into the consideration. Modification to SAX method, FAST-SAX is proposed by M. Fuad where each series in the database is represented by a first-degree polynomial, which is the approximating function for all the time series in the databases [13]. The distances between the time series and their approximating function are computed and stored instead of time series to perform faster than original SAX method.

*5) Frequency Dynamic Interval Class (FDIC):* FDIC is advanced and novel approach of data discretization method which is dynamic in nature presented by A. M. Ahmed et al. [7]. The method consists of two phases. First is dynamic interval class and second is interval merging phase. A dynamic interval phase which tries to generate a number of intervals from data information itself, then generated width intervals by considering the intervals of frequency distribution of the pattern and distance of interval patterns and uses automatically computed threshold based on means distribution to determine the number and the length of intervals.

The second phase merging phase which begins with calculation of distance between known interval class and unknown class. In this phase unknown class intervals that have less distribution, are merged with the nearest point with satisfied minimum threshold. The SV-kNNC algorithm is used to determine the new cut point of unknown class interval. It is dependent on nearest known intervals which have more distribution on data. As a result, when discretization ends, numerical continuous attribute values are transformed into discrete ones based on the dataset characteristics.

This nonparametric and bottom up, method is more advantageous over other methods as the data are grouped together as per their characteristics and frequency of values. But only limitation of this method is it is not incremental.

*B. Supervised Discretization*

There are numbers of supervised discretization methods based on entropy interested readers can check with [2], [3], [14] but here the methods based on clustering [15] and based on Class-Attribute Contingency Coefficient [16] are discussed as they are according to temporal data characteristics and also with class information.

*1) Clustering based Discretization:* This method is taking the advantage of class label and natural grouping together. It is based on clustering, that discretizes the data with the knowledge of both classes and clusters [15]. In this method two clustering algorithms used – the K-means clustering approach with Euclidean distance metric as the similarity measure and the shared nearest neighbour (SNN) clustering algorithm. In the K-means clustering approach the number of clusters is kept approximately equal to the number of classes. Each data instance is assigned to a particular cluster and this is named as the 'pseudo-class'.

Clustering provides the intrinsic grouping of the unlabeled data. Thus, the cluster id captures the interdependencies in the data. So, there are two class features, one provided with data, C, and the other, pseudo-class C'. Then entropy based discretization applied to search for the partition of the value range of a continuous feature so as to minimize the uncertainty of the class variable conditioned on the discretized feature variable. This method results in two intervals and is applied recursively to each subsequent subinterval until the stopping criterion is met.

This method is advantageous because in discretization of continuous variables simultaneously using the class information and cluster based 'pseudo-class' information generally better than that based on the class information alone.

*2) Class Attribute Contingency Coefficient Discretization (CACC) Method:* This is supervised and top-down discretization method based on Class-Attribute Contingency Coefficient by Lee et al. [14]. It calculates CACC value which is used in discretization of continuous data to measure the interdependence between variables.

For each attribute, CACC first finds the maximum and minimum of attributes and then forms a set of all values of each attribute in the ascending order. For all possible interval boundaries and all the midpoints of all the adjacent boundaries in the set are obtained and keeping the maximum CACC value and then partition this attribute accordingly into intervals. The CACC discretization method raise the quality of the generated discretization scheme by extending the idea of contingency coefficient and give better result as considers different attributes together than the consideration of individual attribute, which is the main characteristic of temporal data. But, major limitation is that it is not considering temporal order of data.

From the study of various discretization methods, we derived the following analysis as described in Table 1 and we found that enough research work is done in the area of unsupervised discretization, but very less work is done in the supervised discretization together with incremental approach.

III. SUMMARY AND FUTURE SCOPE

Discretization of data plays an important role in data preprocessing before applying a number of data mining algorithms on the real valued data sets. Here, briefly introduced the need of discretization with the idea and drawbacks of some methods under supervised or unsupervised category for the temporal data. Analysis has been given based on different issues of discretization which shows that still research area is open to consider some issues like nonparametric automatic discretization approach to consider continuous streaming data with the consideration of temporal order.

TABLE I ANALYSIS OF TEMPORAL DATA DISCRETIZATION METHOD

| Category | UnSupervised | | Supervised | | |
|---|---|---|---|---|---|
| Methods / Criteria | EWD, EFD, K-Means | SAX, Extended-SAX, Fast-SAX | FDIC | CACC | Clustering based discretization |
| Supervised or Not | No | No | No | Yes | Yes |
| Consideration of Temporal order | No | Yes | No | No | No |
| Top Down / Bottom Up | Top Down | Top Down | Bottom Up | Top Down | Top Down |
| Parametric | Yes | Yes | No | No | Yes |
| Overlapping solved | No | No | Yes | Yes | No |
| Incremental | No | No | No | Yes | No |

AUTHORS PROFILE

**P. Chaudhari** is M.Tech scholar in Computer Engineering Department at SVNIT, Surat, Gujarat, India.

**D. P. Rana** is Assistant Professor at Computer Engineering Department, S. V. National Institute of Technology, Surat, Gujarat, India and is currently pursuing her PhD degree. Her research interest is in the field of security in web applications, computer architecture, database management system, data mining and web data mining. She is a life member of ISTE and CSI.

**R. G. Mehta** is Associate Professor at Computer Engineering Department, S. V. National Institute of Technology, Surat, Gujarat, India and is currently pursuing her PhD degree. Her research interest is in the field of security in web applications, computer architecture, database management system, data mining and web data mining. She is a life member of ISTE and CSI.

**N. J. Mistry** is Professor at Civil Engineering Department, S. V. National Institute of Technology, Surat, Gujarat-395007, India. He is a member of CES.

**M. M. Raghuwanshi** is working as a principal at Rajiv Gandhi College of Engineering and Research, Nagpur, India. He completed his PhD in Computer Science, 2007, at VNIT, Nagpur, India. He is a member of IEEE, ISTE and CSI.